\documentclass[11pt,a4paper]{article}

\usepackage{subfigure,jcappub,amsfonts,amsmath,amssymb,bm,graphicx}

\usepackage[active]{srcltx}

\title{Magnetic helicity evolution in a neutron star accounting for the Adler-Bell-Jackiw anomaly}

\author[a,b]{Maxim Dvornikov}
\author{and}
\author[a]{Victor B. Semikoz}
\affiliation[a]{Pushkov Institute of Terrestrial Magnetism, Ionosphere
and Radiowave Propagation (IZMIRAN),
108840 Moscow, Troitsk, Russia}
\affiliation[b]{Physics Faculty, National Research Tomsk State University,
36 Lenin Avenue, 634050 Tomsk, Russia}
\emailAdd{maxdvo@izmiran.ru}
\emailAdd{semikoz@yandex.ru}

\abstract{
We analyze the role of the surface terms in the conservation law for the sum of the magnetic helicity density and the chiral imbalance of the charged particle densities. These terms are neglected in the Anomalous MagnetoHydroDynamics (AMHD), where infinite volume is considered typically. We discuss a finite volume system, such as a magnetized neutron star (NS), and study the contribution of the surface terms to the evolution of the magnetic helicity. Accounting for the fast washing out of the chiral imbalance in a nascent NS, we demonstrate that the surface terms contribution can potentially lead to the reconnection of magnetic field lines and subsequent gamma or X-ray bursts observed from magnetars. We derive the additional surface terms originated by the mean spin flux through a volume boundary arising due to macroscopic spin effects in electron-positron plasma. Then, comparing this quantum surface term with the classical one known in standard MHD, we find that the new quantum contribution prevails over classical term for the rigid NS rotation only.
}

\keywords{Magnetohydrodynamics, neutron stars, magnetic fields}

\begin{document}

\maketitle

\section{Introduction}

The assumption of the vanishing masses of charged particles leads to the consideration of the chiral MHD or AMHD~\cite{SonSur09}. Numerous interesting phenomena, like chiral magnetic waves, were predicted in this subject. The review on AMHD is given in ref.~\cite{Zak13}.

The aim of this work is to elucidate which impact of the AMHD remains in the standard MHD when we admit the presence of a non-zero particle mass in magnetized plasma. Let us remind that, in the AMHD, one assumes the zero fermion mass, $m=0$, to support the chiral magnetic effect (the CME) based on the anomalous current \cite{Vilenkin:1980fu} (see below in eqs.~(\ref{anomalous_current}) and~(\ref{CME}), as well as in refs.~\cite{Fukushima:2008xe,BFR,Boyarsky:2015faa}).

In realistic astrophysical and cosmological media, e.g., in a hot plasma of the early universe with $T\gg m_e$, or in the ultrarelativistic degenerate  electron gas of a supernova, where $p_{F_e}\gg m_e$, the particles are massive, $m_e\neq 0$, that leads very soon  to the zero chiral imbalance, $\mu_5=0$, due to the spin-flip through particle collisions~\cite{Dvo16}. 

Accounting for the non-zero mass $m_e\neq 0$ for right and left electrons (positrons), we lose the CME while state here compatibility
of the standard MHD with the AMHD based on the Adler-Bell-Jackiw anomaly, which is valid for massive particles as well~\cite{Iof06}.
This happens under the following additional condition: the losses of magnetic helicity within a domain volume are given by the additional quantum effect, consisting in the nonzero mean spin flux through the surface of that domain. The magnetic helicity evolution includes such a new term which can be competitive with the known helicity losses at the same surface in classical MHD~\cite{Priest}. Let us remind that
the magnetic helicity evolution at a stellar surface is crucial for the reconnection of magnetic field lines there followed by the efficient conversion of the magnetic energy into thermal and kinetic energies of plasma leading, e.g., to the strong electromagnetic emission of highly magnetized compact stars~\cite{ThoLyuKul02}, called magnetars~\cite{MerPonMel15}.

Our work is organized as follows. In section~\ref{sec:MAGNETIZ}, we remind some known results for the spin magnetization in plasma and some formulas describing the electron density in the relativistic degenerate electron gas of a magnetized neutron star (NS). Then, in the main section~\ref{sec:QUANTCORR}, we derive from the Adler-Bell-Jackiw anomaly the surface terms for the modified CME accounting for all spin terms including that originated by the pseudoscalar ($\sim \bar{\psi}\gamma_5\psi$). Here $\gamma_5 = \mathrm{i}\gamma^0\gamma^1\gamma^2\gamma^3$ and $\gamma^\mu = (\gamma^0,\bm{\gamma})$ are the Dirac matrices. In section~\ref{sec:HELDISS}, we derive the new magnetic helicity evolution equation, where, neglecting chiral anomaly $n_\mathrm{R} - n_\mathrm{L}=0$ due to the spin-flip, we find the contribution of the mean spin flux through the boundary of the volume to the dynamics of the magnetic helicity known in classical MHD.
In section~\ref{sec:HELEVOL}, we compare estimates of magnetic helicity losses given by the classical MHD \cite{Priest} and our new (quantum) contribution given by that mean spin flux. In section~\ref{sec:SUMMARY}, we summarize our results. In appendix~\ref{sec:SPINDISTR} we derive the equilibrium spin distribution function for relativistic plasma given by the paramagnetic term in the Landau spectrum (see eq.~(\ref{levels}) below),
and, in appendix~\ref{sec:GAMMA5},  we calculate the mean pseudoscalar $\langle\bar{\psi}\gamma^5\psi\rangle$ using the semiclassical Wentzel-Kramers-Brillouin (WKB) approximation of large Landau numbers, $n\gg 1$.

\section{Mean spin of the electron gas in an external magnetic field\label{sec:MAGNETIZ}}

In this section, we remind the basic properties of plasma in an external magnetic field.

The motion of a relativistic charged particle in a magnetic field obeys the Dirac equation. The energy levels of a $1/2$-spin fermion with the electric charge $q$ in an external magnetic field ${\bf H}=(0,0,H)$ has the form~\cite[pgs.~121--122]{BerLifPit82},
\begin{equation}\label{levels}
  \mathcal{E}(p_z, n,\lambda)  =
  \sqrt{m_e^2 +p_z^2 + |q| H (2n+ 1) - q H \lambda}.
\end{equation}
where $p_z$ is the conserved projection of the fermion momentum along the magnetic field, $n = 0,1,\dots$ is the discrete main quantum number, and $\lambda = \pm 1$ is the eigenvalue of the matrix $\Sigma_z = \gamma^5\gamma^0\gamma^3$, which appears in the squared Dirac equation.\footnote{Strictly speaking, the matrix $\Sigma_z$ does not commute with the Hamiltonian for a moving Dirac particle~\cite[pg.~86]{BerLifPit82}. Nevertheless, it is convenient to use $\lambda$ in the energy levels in eq.~\eqref{levels}.} Negatively charged particles (electrons) should have $q=-e$ and positively charged ones (positrons) possess $q=+e$. Here $e>0$ is the absolute value of the elementary charge.


Let us consider $e^{\pm}$ plasma in the external magnetic field. The main Landau level with $n=0$ was found in ref.~\cite{SV} to contribute the mean spin of electrons and positrons, which has the form,
%
%
%
\begin{align}\label{total}
  \bm{\mathcal{S}} = & 
  \langle \psi_e^\dagger \bm{\Sigma} \psi_e\rangle_{0} =
  - \frac{2e\mathbf{H}}{(2\pi)^2}
  \notag
  \\
  & \times
  \int_0^{\infty}\mathrm{d}p_z
  \left\{
    \frac{1}{\exp[(\sqrt{p_z^2 + m_e^2} -\mu_e)/T] + 1}-
    \frac{1}{\exp[(\sqrt{p_z^2 + m_e^2} +\mu_e)/T] + 1}
  \right\},
\end{align}
where $\psi_e$ is the exact solution of the Dirac equation in the external magnetic field, $\bm{\Sigma} = \gamma^5\gamma^0\bm{\gamma}$,
$\mu_e$ is the chemical potential of electrons, and $T$ is the plasma temperature.
 
In a chiral plasma, $m_e\to 0$, the integral in the last line gives  the chemical potential $\mu_e$ independently of the temperature. Thus the mean spin in a relativistic $e^{\pm}$ plasma 
reads
\begin{equation}\label{chiral-limit}
  \bm{\mathcal{S}} = -
  \frac{e\mu_e{\bf H}}{2\pi^2}= -
  \left(
    \frac{\alpha_\mathrm{em}}{\pi}
  \right)
  \left(
    \frac{2\mu_e}{e}
  \right)
  {\bf H},
\end{equation}
where $\alpha_\mathrm{em}=e^2/4\pi\approx 1/137$ is the fine structure constant.
In a non-relativistic degenerate electron gas, e.g., in metal plasma, substituting in eq.~(\ref{total}) chemical potential $\mu_e=m_e + p^2_{F_e}/2m_e$, one gets the paramagnetic magnetization term $\mathbf{M}=-\mu_\mathrm{B}\bm{\mathcal{S}}$,\footnote{Here we take into account the fact that the magnetic moment of a electron is negative: $-\mu_\mathrm{B}$.}
\begin{equation}\label{NRplasma}
  \frac{\mathbf{M}}{\mu_\mathrm{B}}=\frac{ep_{F_e}}{2\pi^2}{\bf H},
\end{equation}
where $\mu_\mathrm{B}=e/2m_e>0$ is the Bohr magneton.
It means that the static paramagnetic susceptibility  equals to the known value $\chi=\alpha_\mathrm{em}v_{F_e}/\pi\ll 1$ where  $v_{F_e}=p_{F_e}/m_e$ is the Fermi velocity, resulting in the standard definitions $\mathbf{M}=\chi \mathbf{H}$ and $\mathbf{B}=\mu \mathbf{H}= (1 + \chi ) \mathbf{H}$; cf. ref.~\cite[eq.~(59.5) on pg.~173]{LanLif80}. Here $\mu$ is the magnetic permeability of the electron gas, and owing to the fact that $\chi\ll 1$, the approximation ${\bf B}\approx {\bf H}$ is valid with a good accuracy. 

Note that, in a degenerate electron gas (both ultra-relativistic, as in eq.~(\ref{chiral-limit}), and non-relativistic, as in eq.~(\ref{NRplasma})),  the mean spin is produced only by electrons populating the main Landau level $n=0$,
\begin{equation}\label{mainLandau}
  \bm{\mathcal{S}}= - n_{0}\hat{\bf n}_\mathrm{B},
  \quad
  \hat{\bf n}_\mathrm{B}=\frac{\bf B}{B},
  \quad
  n_0=\frac{eBp_{F_e}}{2\pi^2},
\end{equation}
where such number density $n_0$ is a part of the total electron density in the degenerate electron gas \cite{Nunokawa:1997dp}:
\begin{equation}\label{totaldensity}
  n_e=\frac{eBp_{F_e}}{2\pi^2} + \sum_{n=1}^{n_\mathrm{max}}
  \frac{2eB\sqrt{p_{F_e}^2 - 2eBn}}{2\pi^2}.
\end{equation}
Here the summation goes up to a maximum value $n_\mathrm{max}=[p_{F_e}^2/(2eB)]$, the integer part of $p_{F_e}^2/(2eB)$.

In the strong magnetic field limit, $2eB>p_{F_e}^2$, the sum in eq.~(\ref{totaldensity}) vanishes and all electrons populate the main Landau level, $n_e=n_0$. Then, if we consider massless electrons, the complete electric current should be the anomalous current which drives CME~\cite{Vilenkin:1980fu},
\begin{equation}\label{anomalous_current}
{\bf j}=\alpha {\bf B}=\frac{\alpha_\mathrm{em}\mu_5}{\pi}{\bf B},~~~\mu_5=(\mu_\mathrm{R} - \mu_\mathrm{L})/2.
\end{equation}
One can see that the current in eq.~\eqref{anomalous_current} generates the forceless magnetic field, ${\bf j}\times {\bf B}=0$, for which  3-D solution ${\bf B}(r,\theta,\phi)$ to the equation $\nabla\times {\bf B}=\alpha {\bf B}$ is well-known; cf. ref.~\cite{Chandra}. 

Note that a realistic assumption $p_{F_e}\sim 100~{\rm MeV}$ within the core of NS would be problematic in the case $eB>p_{F_e}^2/2$ since the corresponding super-strong magnetic field $B> 10^4 B_\mathrm{cr}/2\sim 2.2\times 10^{17}\,{\rm G}$, where $B_\mathrm{cr} = m_e^2/e = 4.4\times10^{13}\,\text{G}$, that provides the condition $n_e=n_0$, is quite great. 

Nevertheless, the assumption of a moderately strong  magnetic field,
$m_e^2\ll 2eB\ll p_{F_e}^2$, valid for charged components within NS obeying $p_{F_e}=p_{F_p}\sim 100~{\rm MeV}$ in a electroneutral plasma, $n_e=n_p$, would be more realistic, in particular, for magnetars ($B\sim 10^{15}~{\rm G}$). Under such conditions, the sum for Landau levels $1\leq n\leq n_\mathrm{max}$ in eq.~(\ref{totaldensity}) contributes to the complete current significantly more than the anomalous one. Hence the usual transversal components ${\bf B}\perp {\bf j}$ prevail and Lorentz force exists. It means that the CME is negligible since density $n_0$ at the main Landau level is a small correction\footnote{In what follows such a small correction leads to the most important term for the magnetic helicity dissipation in the case of the rigid rotation of NS.} to the total one~\cite{Nunokawa:1997dp},
\begin{equation}\label{real}
n_e\approx \frac{p_{F_e}^3}{3\pi^2}\left[1 + \frac{3eB}{2p_{F_e}^2}\right].
\end{equation}
If we consider the correction to the electron number density from the magnetic field in eq.~\eqref{real}, the chemical potential in eq.~\eqref{total} becomes dependent on the magnetic field.

\section{The quantum surface correction to the chiral anomaly in finite volume of a neutron star\label{sec:QUANTCORR}}

In this section, we start with the system of AMHD equations for chiral particles. Then we show how these equations are modified when particle mass is accounted for. In particular, we study the contribution of the particle mass to the Adler-Bell-Jackiw anomaly.

The full system of AMHD equations reads~\cite{PavLeiSig17},
\begin{align}
  \rho
  \left[
    \frac{\partial \mathbf{v}}{\partial t} + (\mathbf{v} \nabla) \mathbf{v} -
    \nu \nabla^2 \mathbf{v}
  \right] = & 
  - \nabla p + \sigma_\mathrm{cond} [(\mathbf{E} \times \mathbf{B}) +
  (\mathbf{v} \times \mathbf{B}) \times \mathbf{B}],
  \notag
  \\
  \frac{\partial \rho}{\partial t} + \nabla (\rho \mathbf{v}) = & 0,
  \label{eq:hyd}
  \\
  \frac{\partial \mathbf{B}}{\partial t} = & - (\nabla \times \mathbf{E}),
  \label{eq:ind}
  \\
  (\nabla \times \mathbf{B}) = & \sigma_\mathrm{cond} 
  \left[
    \mathbf{E} - \frac{e^2}{2\pi^2\sigma_\mathrm{cond}} \mu_5 \mathbf{B} +
    (\mathbf{v} \times \mathbf{B})
  \right],
  \label{eq:CME}
  \\
  \frac{\mathrm{d} \mu_5}{\mathrm{d}t} = &
  - \frac{e^2}{4\mu_e^2} \frac{\mathrm{d} h}{\mathrm{d}t},
  \label{eq:triangle}
\end{align}
where $\rho$ is the matter density, $\mathbf{v}$ is the fluid velocity, $p$ is the pressure, $\sigma_\mathrm{cond}$ is the electric conductivity, $\mu_5=(\mu_{e\mathrm{R}} - \mu_{e\mathrm{L}})/2$ is the chiral imbalance of right and left electrons, $h$ is the magnetic helicity density, $\mu_e$ is the mean chemical potential of electrons, and $\nu$ is the viscosity coefficient.

Navier-Stokes and the continuity equations are grouped in eq.~\eqref{eq:hyd}. In eq.~\eqref{eq:CME}, the anomalous current owing to the CME is accounted for in the Maxwell equation where the displacement current is neglected. The Faraday equation, resulting from eqs.~\eqref{eq:ind} and~\eqref{eq:CME}, is modified owing to the CME,
\begin{equation}\label{Faraday}
  \frac{\partial \mathbf{B}}{\partial t} =
  \nabla\times (\mathbf{v}\times \mathbf{B}) +
  \frac{1}{\sigma_\mathrm{cond}}\nabla^2\mathbf{B} -
  \frac{e^2}{2\pi^2\sigma_\mathrm{cond}}\nabla\times (\mu_5\mathbf{B}).
\end{equation}
Equation~\eqref{eq:triangle} is a consequence of the Adler-Bell-Jackiw anomaly (see below) for massless particles. Note that we assume that electron gas is degenerate in eq.~\eqref{eq:triangle}. In the following, we shall modify eq.~\eqref{eq:triangle} accounting for the nonzero electron mass.

Let us elucidate which terms are important for the QED anomaly in a finite volume, for instance, in NS.
In particular, we calculate the correction due to the mean massive term $2\mathrm{i}m_e\langle \bar{\psi}\gamma_5\psi\rangle$ to the Adler-Bell-Jackiw anomaly in QED known as the non-conservation of pseudovector current,
\begin{equation}\label{pseudovector}
\frac{\partial }{\partial x^{\mu}}\bar{\psi}\gamma^{\mu}\gamma^5\psi=\frac{\partial j_\mathrm{R}^{\mu}}{\partial x^{\mu}} - \frac{\partial j_\mathrm{L}^{\mu}}{\partial x^{\mu}} = 2\mathrm{i}m_e\bar{\psi}\gamma_5\psi + \frac{2e^2}{16\pi^2}F_{\mu\nu}\tilde{F}^{\mu\nu},
\end{equation}
where right and left currents do not persist separately even in the massless case due to the Abelian (triangle) anomaly in the presence of elecromagnetic fields:
\begin{equation}\label{rightleft}
\frac{\partial j_\mathrm{R,L}^{\mu}}{\partial x^{\mu}}=\pm \mathrm{i}m_e\bar{\psi}\gamma_5\psi \pm \frac{e^2}{16\pi^2}F_{\mu\nu}\tilde{F}^{\mu\nu}.
\end{equation}
Using the fact that $\gamma^0\bm{\gamma}\gamma^5 = \bm{\Sigma}$ and averaging the left hand side in eq.~(\ref{pseudovector}), one gets
\begin{equation}\label{lhs}
  \int\frac{\mathrm{d}^3x}{V}\frac{\partial }{\partial x^{\mu}}
  \langle\bar{\psi}\gamma^{\mu}\gamma^5\psi\rangle =
  \frac{{\rm d}}{{\rm d}t}(n_\mathrm{R} - n_\mathrm{L}) +
  \frac{1}{V}\oint \mathrm{d}^2S(\bm{\mathcal{S}}\cdot{\bf n}),
\end{equation}
where $\bm{\mathcal{S}}$ is the mean spin given for massless particles by eq.~(\ref{chiral-limit}).
Averaging the right hand side in eq.~(\ref{pseudovector}), we obtain
\begin{equation}\label{rhs}
-\frac{1}{V}\oint_S (\bm{\mathcal{S}}_{5}\cdot {\bf n}) \mathrm{d}^2S + \frac{2\alpha_\mathrm{em}}{\pi}\int \frac{\mathrm{d}^3x}{V}({\bf E}\cdot{\bf B}),
\end{equation}
where the first term $\sim \bm{\mathcal{S}}_5$ is stipulated by the mean pseudoscalar,
\begin{equation}\label{mean5}
2\mathrm{i}m_e\int \frac{\mathrm{d}^3x}{V}\langle\bar{\psi}\gamma_5\psi\rangle=
-\int\frac{\mathrm{d}^3x}{V}(\nabla\cdot\bm{\mathcal{S}}_{5}({\bf x},t)),
\end{equation}
while the second term corresponds to the magnetic helicity dissipation given by the standard (classical) MHD \cite{Priest},
\begin{equation}\label{standard1}
 2\int \frac{\mathrm{d}^3x}{V}({\bf E}\cdot{\bf B})= -\frac{{\rm d}h}{{\rm d}t} - \oint({\bf n}\cdot\left[{\bf B}A_0 +{\bf E}\times {\bf A}\right])\frac{\mathrm{d}^2S}{V}.
\end{equation}
The term in eq.~(\ref{mean5}) arises due to the spin distribution functions in a weakly inhomogeneous $e^{\pm}$ plasma (see derivation in appendix~\ref{sec:GAMMA5}):
\begin{equation}\label{magnetization}
\bm{\mathcal{S}}_{5}({\bf x},t)=-\int\frac{\mathrm{d}^3p}{\gamma(2\pi)^3}\left({\bf S}^{(e)}({\bf p},{\bf x},t) - {\bf S}^{(\bar{e})}({\bf p},{\bf x},t)\right)\left(\frac{2}{3} + \frac{1}{3\gamma}\right).
\end{equation}
Here the equilibrium part of the total Wigner's spin  distribution function for electrons, ${\bf S}^{(e)}({\bf p},{\bf x},t)={\bf S}^{(e)}_\mathrm{eq}(\varepsilon_p,{\bf x},t) + \delta {\bf S}^{(e)}({\bf p},{\bf x},t)$, originated by the paramagnetic contribution in a low inhomogeneous magnetic field  is well-known (see appendix~\ref{sec:SPINDISTR} and ref.~\cite{Silin}):
\begin{equation}\label{spin_0}
  {\bf S}^{(e)}_\mathrm{eq}(\varepsilon_p,{\bf x},t)=
  \frac{\mu_\mathrm{B} {\bf H}({\bf x},t)}{\gamma}
  \frac{{\rm d}f^{(e)}_\mathrm{eq}(\varepsilon_p)}{{\rm d}\varepsilon_p},
  \quad
  \varepsilon_p=\sqrt{p^2 + m_e^2}=\gamma m_e.
\end{equation}
Combining eqs.~(\ref{lhs}) and~(\ref{rhs}), as well as accounting for the standard eq.~(\ref{standard1}) for the magnetic helicity density $h(t)=V^{-1}\int \mathrm{d}^3x({\bf A}\cdot{\bf B})$, we derive the master equation,
\begin{align}\label{new_law}
 \frac{{\rm d}}{{\rm d}t}
  \left(
    n_\mathrm{R} - n_\mathrm{L} + \frac{\alpha_\mathrm{em}}{\pi}h
  \right) =
  - \frac{\alpha_\mathrm{em}}{\pi V}\oint_S([{\bf E}\times {\bf A} + A_0{\bf B}]\cdot{\bf n})\mathrm{d}^2S - 
  \oint_S
  ([ \bm{\mathcal{S}} + \bm{\mathcal{S}}_{5}]\cdot {\bf n})
  \frac{\mathrm{d}^2S}{V}.
\end{align}
Neglecting the surface terms in a chiral plasma of massless particles, one gets the conservation law for the sum of the imbalance of right and left particle densities, $n_\mathrm{R} - n_\mathrm{L}$, and the magnetic helicity density $h$ (multiplied by $\alpha_\mathrm{em}/\pi$),
\begin{equation}\label{CME}
\frac{{\rm d}}{{\rm d}t}\left(n_\mathrm{R} - n_\mathrm{L} +\frac{\alpha_\mathrm{em}}{\pi}h\right)=0.
\end{equation}
The relation in eq.~(\ref{CME}) is essential for the CME since it describes the dynamics of the chiral imbalance ($\sim {\rm d}\mu_5/{\rm d}t$) in a magnetized medium~\cite{Fukushima:2008xe,BFR}: a decrease of such imbalance
leads to generation of the magnetic helicity, and vice versa. 

Eq.~(\ref{new_law}) includes the known classical surface term, containing the electromagnetic fields only, and a new quantum correction originated by the sum of spin terms, $\bm{\mathcal{S}}_\mathrm{eff}=\bm{\mathcal{S}} + \bm{\mathcal{S}}_5$. We will study the latter term below.

 Let us stress that the equilibrium spin distribution function for the positron gas has the same positive sign as in the case of electrons (\ref{spin_0}),
\begin{equation}\label{spin_00}
{\bf S}^{(\bar{e})}_\mathrm{eq}(\varepsilon_p,{\bf x},t)= \frac{\mu_\mathrm{B} {\bf H}({\bf x},t)}{\gamma}\frac{{\rm d}f^{(\bar{e})}_\mathrm{eq}(\varepsilon_p)}{{\rm d}\varepsilon_p}.
\end{equation}
Thus the pseudovector in eq.~(\ref{magnetization}) is given by the difference of particle and antiparticle contributions, owing to the operator permutation $\hat{d}\hat{d}^+\to - \hat{d}^+\hat{d}$ when deriving eq.~(\ref{magnetization}),
\begin{equation}\label{magnetization2}
  \bm{\mathcal{S}}_{5}({\bf x},t)= - \frac{\mu_\mathrm{B} m_e^2{\bf H}({\bf x},t)}{2\pi^2}\int \frac{p^2\mathrm{d}p}{\varepsilon_p^2}\left(\frac{{\rm d}f^{(e)}_\mathrm{eq}}{{\rm d}\varepsilon_p} - \frac{{\rm d}f^{(\bar{e})}_\mathrm{eq}}{{\rm d}\varepsilon_p}\right)\left(\frac{2}{3} + \frac{1}{3\gamma}\right).
\end{equation}
Integrating by parts in eq.~(\ref{magnetization2}) and separating the Lande factor $g_s=2$ from the Fermi distributions $f^{(e,\bar{e})}(\varepsilon_p)=g_s[\exp (\varepsilon_p \mp \mu_e)/T +1]^{-1}$, we can rewrite the effective mean spin $\bm{\mathcal{S}}_\mathrm{eff}=\bm{\mathcal{S}} + \bm{\mathcal{S}}_5$ entering the surface term in the master  eq.~(\ref{new_law}), as
\begin{align}\label{quantum_correction}
  \bm{\mathcal{S}}_\mathrm{eff} = &
  - \frac{e{\bf H}}{2\pi^2}\int_0^{\infty}\mathrm{d}p
  \left[
    1 -
    \frac{1}{3\gamma^2}
    \left(
      \frac{2}{\gamma} + \frac{2}{\gamma^2} - 1
    \right)
  \right]
  \notag
  \\
  & \times
  \left(
    \frac{1}{\exp [(\varepsilon_p - \mu_e)/T] +1} -
    \frac{1}{\exp [(\varepsilon_p + \mu_e)/T] +1}
  \right).
\end{align}
In a non-relativistic plasma, $\varepsilon_p=\gamma m_e\approx m_e$. Thus the combined spin effect practically vanishes, $\bm{\mathcal{S}}_\mathrm{eff}=\bm{\mathcal{S}} + \bm{\mathcal{S}}_5  \to 0$, since both spin terms compensate each other.  For example, in a degenerate electron gas, where the positron contribution is absent, one obtains
$\bm{\mathcal{S}} = - \bm{\mathcal{S}}_5 = - 2 e m_e v_{F_e}{\bf B}/4\pi^2$, where $v_{F_e}=p_{F_e}/m_e\ll 1$, see in eq.~(\ref{NRplasma}). The first correction to the last equality arises due to decomposition $\gamma^{-1}\approx 1 - v^2/2$. It turns out to be rather small, $\bm{\mathcal{S}}_\mathrm{eff} = e m_e v_{F_e}^3{\bf B}/3\pi^2$. This correction is applicable in the region of weak magnetic fields, $m_e^2\gg p_{F_e}^2\gg eB$ where the WKB approximation with large Landau numbers, $n\gg 1$, used for calculation of $\bm{\mathcal{S}}_5$, remains correct.

In the case of ultrarelativistic chiral plasma ($\gamma\gg 1$) one obtains from eq.~(\ref{quantum_correction}) $\bm{\mathcal{S}}_\mathrm{eff} \approx \bm{\mathcal{S}}= - e\mu_e(r,\theta) {\bf B}/2\pi^2$, see in eq.~(\ref{chiral-limit}). Hence the magnetization effect is great under the condition $\mu_e( r,\theta)\gg m_e$, which can be implemented, e.g., in the core of NS. Here we input intentionally an inhomogeneous chemical potential $\mu_e(r,\theta)$  that corresponds to a real spherically symmetric electron density profile in NS~\cite{Lattimer}, $n_e(r)=n_\mathrm{core}Y_e[1 - r^2/R_\mathrm{NS}^2]$, where $Y_e=0.04$ is the electron abundance and $n_\mathrm{core}\simeq 10^{38}\,\text{cm}^{-3}$ is the central (neutron) density.

For instance, in the axially symmetric (rotating) star the anisotropy for the chemical potential results from the magnetic field correction proportional to the total magnetic field entering Landau levels\footnote{We consider slightly non-uniform magnetic field, i.e. it is uniform at microscopic scales that are less than the mean distance between particles in medium, $L< (n_e)^{-1/3}$. This requirement is necessary to find the energy levels in eq.~\eqref{levels}.
Nevertheless ${\bf B}$ is considered to be non-uniform at macroscopic scales comparable with $R_\mathrm{NS}$.}, $ B(r,\theta)=\sqrt{B_{p}^2(r,\theta) + B_{t}^2(r,\theta)}$, where $B_{p,t}$ are the poloidal and toroidal components of the magnetic field (see section~\ref{sec:HELEVOL}), and derived from inverted eq.~(\ref{real}),\footnote{We mean the cubic eq.~(\ref{real}) rewritten as $\mu^3_e + 3eB\mu_e/2 - 3\pi^2n_e=0$, where $\mu_e\equiv p_{F_e}$. The parameter $D=q^2 + p^3= (eB)^3/8 + 9\pi^4n_e^2/4>0$ is positive for Cartan solution. Hence there are one real root of this equation shown in eq.~(\ref{anisotrop})  and two complex self-adjoint ones.}
\begin{equation}\label{anisotrop}
\mu_e(r,\theta)=(3\pi^2n_e(r))^{1/3}\left[1 - \frac{eB(r,\theta)}{2(3\pi^2n_e(r))^{2/3}}\right].\end{equation}
 Otherwise,  the surface integral
\begin{equation}\label{profit}
  \oint_S(\bm{\mathcal{S}}\cdot{\bf n})\frac{\mathrm{d}^2S}{V}=
  - \frac{e}{2\pi^2 V}
  \oint_S\mu_e(r,\theta)({\bf B}\cdot{\bf n})\mathrm{d}^2S,
\end{equation}
remaining in the master eq.~(\ref{new_law}) for NS, should vanish for the uniform spherically symmetric $\mu_e(r)_{r=R}=\text{const}$ due to Gauss theorem, $\int \mathrm{d}^3x (\nabla\cdot {\bf B})=\oint ({\bf B}\cdot{\bf n})\mathrm{d}^2S=0$.

\section{Magnetic helicity dissipation as the mean spin flux through a domain boundary\label{sec:HELDISS}}

For the compatibility of the magnetic helicity density evolution in eq.~(\ref{standard1}), 
and statistically averaged Adler-Bell-Jackiw anomaly given in a chiral medium  by eq.~(\ref{new_law}), let us consider the realistic situation when the chiral imbalance  vanishes in relativistic plasmas accounting for the non-zero electron mass, $m_e\neq 0$, due to the spin-flip , $n_\mathrm{R} - n_\mathrm{L}\to 0$. As shown in refs.~\cite{Dvornikov:2014uza,Dvo16}, this situation happens in the core of a nascent NS very soon, during $\sim 10^{-12}\,\text{s}$, even for an initial positive difference $n_\mathrm{R} (t_0) - n_\mathrm{L}(t_0) > 0$ arising in a supernova's progenitor of that NS at the initial moment $t_0$ owing to the direct Urca-process, $p+ e_\mathrm{L}^-\to n + \nu_{e\mathrm{L}}$. The similar decrease of the chiral imbalance $2\mu_5=\mu_\mathrm{R} - \mu_\mathrm{L}\to 0$ due to the spin-flip down to the temperature $T=10\,{\rm MeV}$ in the cooling universe, was found in ref.~\cite{BFR}. 

Then, at time $t\gg t_0$, we should modify the standard MHD eq.~(\ref{standard1}) due to eq~(\ref{new_law}), accounting for
the magnetic flux through volume surface weighted by the nonuniform chemical potential $\mu_e(r,\theta)$ in eq.~(\ref{anisotrop}) that enters the quantum (magnetization) term  in eq.~(\ref{profit}),
\begin{align}\label{dissipation}
  \oint  \mathrm{d}^2S
  \left(
    \bm{\mathcal{S}}_\mathrm{eff}\cdot{\bf n}
  \right)
  \approx
  & 
  - \frac{e}{2\pi^2}
  \oint_S\mu_e(r,\theta)({\bf B}\cdot {\bf n})\mathrm{d}^2S
  \nonumber
  \\
  & =
  \frac{\alpha_\mathrm{em}}{\pi (3\pi^2n_e(R))^{1/3}}
  \oint B(R,\theta)({\bf B}\cdot {\bf n})\mathrm{d}^2S.
\end{align}
Here the non-uniform electron density $n_e (R)=n_\mathrm{core}Y_e(1 - R^2/R^2_{NS})$ should be large enough to obey the inequality $2eB(R,\theta)\ll (3\pi^2n_e (R))^{2/3}$ at the surface with radius $R< R_\mathrm{NS}$ since the decomposition is made over the small parameter $2eB/\mu_e^2\ll 1$ in eq.~(\ref{real}).
Then, we can generalize standard eq.~(\ref{standard1}) due to the  Adler-Bell-Jackiw anomaly accounting for the additional quantum contribution in eq.~(\ref{dissipation}),
\begin{equation}\label{standard3}
\frac{{\rm d}H}{{\rm d}t}= - 2\int_V\mathrm{d}^3x ({\bf E}\cdot{\bf B})-\oint_S
({\bf n}\cdot[A_0{\bf B} + {\bf E}\times {\bf A}])\mathrm{d}^2S + \frac{1}{\mu_e(R)}\oint_S B(R,\theta)B_r(R,\theta)\mathrm{d}^2S,
\end{equation}
where $B(R,\theta)=\sqrt{B_r^2 + B_{\theta}^2 + B_{\varphi}^2}$ is the total magnetic field strength entering Landau levels and we put a short notation $\mu_e(R)=[3\pi^2n_e(R)]^{1/3}$.

There is a statement in ref.~\cite[below eq.~(8.52) in chapter~8]{Priest} that the volume term in the left hand side in eq.~(\ref{standard1}) can be omitted at times much less than the diffusion time $t\ll \tau_\mathrm{D}\sim L^2\sigma_\mathrm{cond}$, so a change of magnetic helicity is determined only by the second term  in eq.~(\ref{standard3}) or the last term in eq.~(\ref{standard1}). The diffusion time turns out to be huge, $\tau_\mathrm{D}\simeq 30\,\text{yr}(L/\text{cm})^2$, due to the great conductivity within NS~\cite{Kelly}, $\sigma_\mathrm{cond} \sim 10^9\,{\rm MeV}$ for $T = 10^8\,\text{K}$. For instance, for the maximal scale $L=R_\mathrm{NS}=10^6\,\text{cm}$ the diffusion time exceeds the age of the universe, $\tau_\mathrm{D} \approx 3 \times 10^{13}\,\text{yr}\gg t_\mathrm{Univ}=1.4\times 10^{10}\,\text{yr}$. Hence for such large scales there is no a reason to take into account both the magnetic helicity diffusion and the quantum term in eq.~(\ref{dissipation}) which both could be essential only at times $t> \tau_\mathrm{D}$. However, at small scales $L\ll R_\mathrm{NS}$ the magnetic helicity diffusion time is less than the age of young magnetars $\sim 10^3\,\text{yr}$, e.g., for $L=1\,\text{cm}$, $\tau_\mathrm{D}\sim 30\,\text{yr}\ll t\sim 10^3\,\text{yr}$. Therefore the quantum contribution to the evolution eq.~(\ref{standard3}) missed in classical approach \cite{Priest}, can be essential for small-scale magnetic fields in NS at times $t>\tau_\mathrm{D}$. 

\section{Evolution of the magnetic helicity in NS\label{sec:HELEVOL}}
 
The study of the magnetic helicity evolution is important for a possible  reconnection of magnetic field lines near the NS surface  happening mostly outside the crust in the NS magnetosphere. This process, in its turn, could explain gamma or X-ray flares observed from magnetars~\cite{ThoLyuKul02,MerPonMel15}. However, it  is interesting to study also how dissipation of the magnetic helicity proceeds inside NS beneath the crust.

Using gauge $A_0=0$ and $(\nabla\cdot{\bf A})=0$ in eq.~(\ref{standard1}), valid in classical MHD, and substituting the Ohm law ${\bf E}= - {\bf v}\times{\bf B} + {\bf j}/\sigma_\mathrm{cond}$,  one can rederive eq.~(8.52) in ref.~\cite{Priest},
\begin{equation}
\frac{{\rm d}H}{{\rm d}t}= - 2\sigma_\mathrm{cond}^{-1}\int {\bf j}\cdot{\bf B}\mathrm{d}^3x -\oint_S[({\bf B}\cdot{\bf A})({\bf v}\cdot{\bf n}) - ({\bf v}\cdot{\bf A})({\bf B}\cdot{\bf n})]\mathrm{d}^2S, 
\end{equation}
that is generalized in eq.~(\ref{standard3}) in the same gauge,
\begin{eqnarray}\label{surface_new}
  \frac{{\rm d}H}{{\rm d}t}= &&- 2\sigma_\mathrm{cond}^{-1}\int {\bf j}\cdot{\bf B}\mathrm{d}^3x
  -
  \oint_S[({\bf B}\cdot{\bf A})({\bf v}\cdot{\bf n}) -
  ({\bf v}\cdot{\bf A})({\bf B}\cdot{\bf n})]\mathrm{d}^2S \nonumber\\&&+ 
    \frac{1}{\mu_e(R)}
  \oint_S B(R,\theta)({\bf B}\cdot {\bf n})\mathrm{d}^2S .
\end{eqnarray}
The last term in the right hand side of eq.~\eqref{surface_new} is originated by the quantum effect of the mean spin flux through surface $R<R_{NS}$ and should be essential at small scales, as shown below. Note that for a closed volume, where $({\bf v}\cdot{\bf n})=0$ and $({\bf B}\cdot {\bf n})=0$, the magnetic helicity in eq.~(\ref{surface_new}) is conserved, ${\rm d}H/{\rm d}t=0$, when one neglects the volume losses for an ideal plasma, $\sigma_{\rm cond}\to \infty$. If magnetic field penetrates the domain volume, i.e. $({\bf B}\cdot {\bf n})\neq 0$, both the quantum contribution and the classical surface term differ from zero.

\begin{figure}
  \centering
  \includegraphics[scale=0.5]{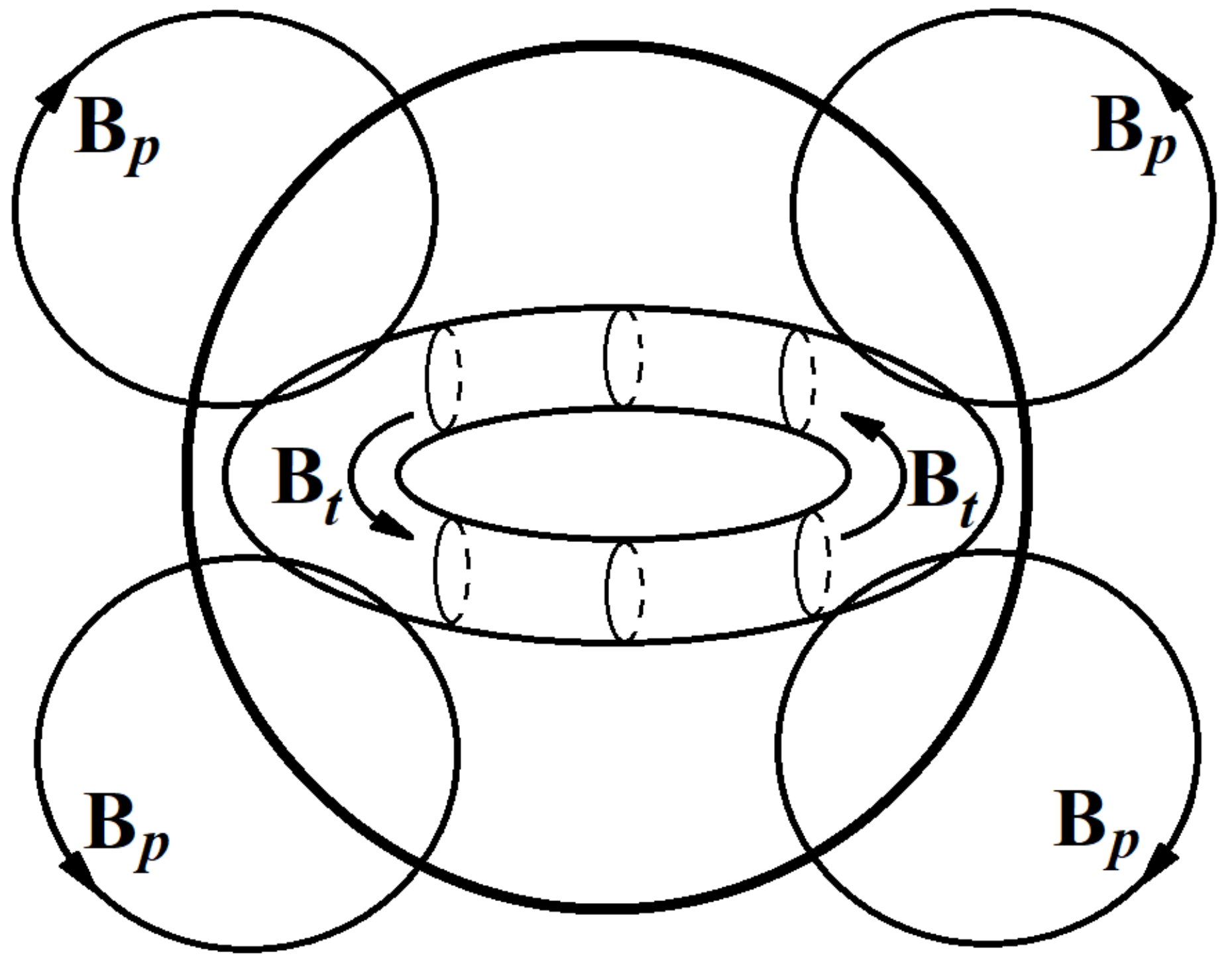}
  \caption{The schematic illustration of the magnetic field configuration given
  in eq.~\eqref{model}. The toroidal component is
  $\mathbf{B}_t = B_{\varphi}(r)\cos \theta {\bf e}_{\varphi}$ and the poloidal field reads
  $\mathbf{B}_p = B_{p}(r)[\cos 2\theta {\bf e}_r + \sin 2\theta {\bf e}_{\theta}]$.  
  \label{fig:magfields}}
\end{figure}

Let us estimate the new quantum contribution, i.e. the last term in eq.~(\ref{surface_new}) that does not  explicitly depend 
on NS rotation, for the axially symmetric magnetic field consisting of the quadrupole poloidal and the toroidal fields~\cite{MSS},
\begin{equation}\label{model}
  {\bf B}(r,\theta)= B_{p}(r)[\cos 2\theta {\bf e}_r + \sin 2\theta {\bf e}_{\theta}] +
  B_{\varphi}(r)\cos \theta {\bf e}_{\varphi},
\end{equation}
where the angle $\theta$ is measured from the equator of NS, which corresponds to $\theta=0$. Note that the magnetic fields components are non-vanishing at the equator. The structure of the magnetic field in eq.~\eqref{model} is schematically illustrated in figure~\ref{fig:magfields}.

The factors $({\bf B}\cdot{\bf n})=B_r=B_p\cos 2\theta$ at the spherical surface where ${\bf n}={\bf e}_r$, and $B(R,\theta)=\sqrt{B_p^2 + B_{\varphi}^2\cos^2\theta}$ should be substituted
into the last term in eq.~(\ref{surface_new}). Then, one obtains such a surface term in the form,
\begin{align}\label{quantum}
  \left(
    \frac{{\rm d}H}{{\rm d}t}
  \right)_\mathrm{quantum} = &
  \frac{1}{\mu_e(R)}
  \oint_S B(R,\theta)({\bf B}\cdot {\bf n})\mathrm{d}^2S
  \nonumber
  \\
  & =-
  \left(
    \frac{2\pi R^2B_p}{\mu_e}
  \right)
  \int_{-1}^1(2x^2 - 1)\sqrt{B_p^2 + B_{\varphi}^2x^2}\mathrm{d}x
  \nonumber
  \\
  & = 
  \left(
    \frac{2\pi R^2B_pB_{\varphi}}{\mu_e}
  \right)
  \times A
  \sim
  \frac{B^2R^2}{\mu_e},
\end{align}
where $\mathrm{d}^2S=2\pi R^2\sin \theta \mathrm{d}\theta$ and the factor $A$,
\begin{align}\label{factor}
  A(B_p/B_t) = & -\frac{1}{2}\left(\frac{B_p}{B_{\varphi}}\right)^2\sqrt{1 + \left(\frac{B_p}{B_{\varphi}}\right)^2} +\left(\frac{B_p}{B_{\varphi}}\right)^2\left[1 + \frac{1}{2}\left(\frac{B_p}{B_{\varphi}}\right)^2\right]
  \nonumber
  \\
  & \times
  \ln \left(\frac{B_{\varphi}}{B_p} + \frac{B_{\varphi}}{B_p}\sqrt{1 + \left(\frac{B_p}{B_{\varphi}}\right)^2}\right),
\end{align}
depends on the ratio $B_p/B_t$. For example, for $B_p\sim B_{\varphi}$ one gets $A\approx -0.05$ and last estimate in eq. (\ref{quantum}) is valid, $| ( \mathrm{d}H/ \mathrm{d}t)_\mathrm{quantum}| \sim B^2R^2/\mu_e$.

The factor $A$ can be simplified in the situation, when $B_{\varphi}\gg B_{p}$,
\begin{equation}\label{factor2}
A=A(B_p/B_t)=\left(\frac{B_p}{B_{\varphi}}\right)^2\left[\ln \left(\frac{2B_{\varphi}}{B_p}\right) - \frac{1}{2}\right].
\end{equation}
In the opposite case, when $B_p\gg B_{\varphi}$, one gets
\begin{equation}\label{factor3}
\left(\frac{{\rm d}H}{{\rm d}t}\right)_\mathrm{quantum}= \frac{4\pi R^2B_p^2}{3\mu_e},
\end{equation}
which can be obtained from eq.~(\ref{quantum}).

The result in eq.~(\ref{quantum}) should be compared with the classical surface term,
\begin{eqnarray}\label{classic}
  \left(
    \frac{{\rm d}H}{{\rm d}t}
  \right)_\mathrm{classic}=&&-\oint_S[({\bf B}\cdot{\bf A})({\bf v}\cdot{\bf n}) - ({\bf v}\cdot{\bf A})({\bf B}\cdot{\bf n})]\mathrm{d}^2S 
\nonumber\\&&
= + \oint (vA_{\varphi})({\bf B}\cdot{\bf n})\mathrm{d}^2S \sim  - B^2_pR^3\langle v\rangle,  
\end{eqnarray}
where we choose the azimuthal rotation velocity ${\bf v}=v{\bf e}_{\varphi}$, i.e. the first term in eq.~(\ref{classic}) vanishes, $({\bf v}\cdot{\bf n})=0$. Then the rotation velocity $v=\Omega R\cos \theta$ and $\langle v\rangle \approx \Omega R$ corresponds to the mean velocity after integration in eq.~(\ref{classic}) over the polar angle. Note that for the poloidal component in eq.~(\ref{model}) $B_{\theta}= - r^{-1}\partial_r(rA_{\varphi})=B_p\sin 2\theta$, one can estimate the azimuthal potential $A_{\varphi}\sim - B_pR\sin 2\theta $
at the sphere $R< R_\mathrm{NS}$.

Comparing the quantum and classical surface terms in the total sum
\begin{equation}
  \left(
    \frac{{\rm d}H}{{\rm d}t}
  \right)=  \left(
    \frac{{\rm d}H}{{\rm d}t}
  \right)_\mathrm{quantum} +   \left(
    \frac{{\rm d}H}{{\rm d}t}
  \right)_\mathrm{classic},
\end{equation}
we find that the first term is bigger only in the co-rotational reference frame, $\langle v\rangle - \Omega R= \delta v\ll 1$, namely for $R< \mu_e^{-1}/\delta v$. Substituting $R=10^5\,\text{cm}$ and $\mu_e^{-1}=2\times 10^{-13}\,\text{cm}$ for $\mu_e=100\,{\rm MeV}$, one gets $\delta v < 10^{-18}$. It correspond to the rigid rotation, $\partial_{\theta}\Omega=\partial_r\Omega=0$, i.e. $\Omega=\text{const}$. In this case, $\langle v\rangle - \Omega R\to 0$. Note that the superfluid neutron component in NS has some deviations from the rigid rotation~\cite{YakPet04}, $\delta v_n\neq 0$, contrary to the proton one, $\delta v_p=0$. It means that, in this situation, our assumption on the absence of any differential rotation should be problematic. 

The result in eq.~(\ref{quantum}), irrespective of NS rotation, can be interpreted as an intertwining of the two thin magnetic tubes with the small base areas $S_p=\pi R_p^2$ and $S_t=\pi R_t^2$ placed at the sphere with radius $R\gg R_{p,t}$~\cite{Priest},
\begin{equation}\label{quantum2}
  \left(
    \frac{{\rm d}H}{{\rm d}t}
  \right)_\mathrm{quantum} =\dot{\theta}_\mathrm{pt}F_pF_t,
\end{equation}
where $F_p=B_pS_p$ and $F_t=B_tS_t$ are magnetic fluxes which tear off the two different toroids (from quadrupole poloidal and toroidal components in eq. (\ref{model})) and then penetrate the spherical surface $R<R_\mathrm{NS}$ floating up. The parameter
\begin{equation}\label{velocity}
\dot{\theta}_\mathrm{pt}=\frac{A}{\pi^2}\times \left(\frac{R}{R_p}\right)^2\left(\frac{1}{R_t^2\mu_e}\right)
\end{equation}
gives the angular velocity with which the magnetic loop bases are twisting one around other causing the interlacing of flux tubes. For example, one can estimate this parameter as $\dot{\theta}_\mathrm{pt}\sim 10^7 \times (6A/\pi^2)\,\text{s}^{-1}$ for $R_p=R_t= 1\,\text{cm}$, $R=10^5\,\text{cm}$. Thus during the time $t\sim 3\times 10^{-6}\,\text{s}\gg 10^{-12}\,\text{s}$ \footnote{We substitute for estimates $A=-0.05$ for $B_p=B_t$ in eq. (\ref{factor}) when already $n_\mathrm{R}=n_\mathrm{L}$. It means that the CME is irrelevant in NS, $\mu_5=0$.}, the magnetic helicity evolution (\ref{velocity}) leads to a flux tangling with the linkage (topology) number $L_\mathrm{pt}=\int \dot{\theta}_\mathrm{pt}dt=L_{12}\sim 1$ in the famous Gauss formula $H= 2L_{12}F_1F_2$ where $L_{12}=L_\mathrm{pt}= 1$ is conserved afterwards; cf. ref.~\cite{Priest}.

\section{Discussion and outline\label{sec:SUMMARY}}

Statistically averaging Adler-Bell-Jackiw anomaly in eq.~(\ref{pseudovector}), we derived our master eq.~(\ref{new_law}) with the new quantum surface term in eq.~(\ref{quantum_correction}) given by the spin effects in plasma. Such a new term becomes important within a finite volume of a dense NS for the magnetic helicity evolution at the spherical surface around that volume.

The magnetic helicity evolution itself could potentially lead to the reconnection of the magnetic field lines at the surface and causing flares from outer boundary of a star happening rather in its magnetosphere. In the present work we did not solve such a problem trying to find only how strong can be new magnetization effect deeply within NS core where our approximations are valid. For that task we should consider small base areas $L^2=(R\Delta \theta)^2\ll R^2$ at the chosen surface with radius $R< R_\mathrm{NS}$ for corresponding thin magnetic tubes intersecting such a surface for both components $B_{p,\varphi}$ in eq. (\ref{model}), since both the magnetic helicity diffusion $-2\int_V \mathrm{d}^3x ({\bf E}\cdot{\bf B})$ and the quantum term in eq.~(\ref{dissipation}) could be important only at small scales, $L\ll R$, when evolution time exceeds a big diffusion time for the high conductivity in NS, $t > \tau_\mathrm{D}=L^2\sigma_\mathrm{cond}$. Note that we do not compare our new term with the diffusion losses considering above an ideal plasma in the limit $\sigma_\mathrm{cond}\to \infty$.

We find that in magnetars with strong magnetic fields $B\sim 10^{15}\,{\rm G}$, for which deeply within core with ultrarelativistic electrons, $p_{F_e}=100\,{\rm MeV}$, the inequalities $m_e^2\ll eB\ll p_{F_e}^2$  are fulfilled, the CME contribution occurs small due to a small population of electrons at the main Landau level $n=0$; cf. eq.~(\ref{real}). Nevertheless, namely these electrons provide the magnetic helicity diffusion through the new quantum contribution in evolution eq.~(\ref{standard3}). 

Note that the WKB approximation $n\gg 1$ in eq.~(\ref{real}), when paramagnetic (spin) contribution is a small correction in the Landau spectrum eq.~(\ref{levels}), simplifies derivation of the pseudoscalar term $2\mathrm{i}m_e\langle \bar{\psi}\gamma_5\psi\rangle=\mu_\mathrm{B}^{-1}(\nabla\cdot \bm{\mathcal{S}}_5)$ made in appendix~\ref{sec:GAMMA5}. 

On the other hand, in the NS crust, where degenerate electrons become non-relativistic,
$eB\gg m_e^2\gg p_{F_e}^2$, and populate the main Landau level $n=0$ only, the WKB approximation is not allowed. In such a case, we plan to calculate anew the pseudovector $\bm{\mathcal{S}}_5$ which is expected to be comparable with the standard magnetization
eq.~(\ref{NRplasma}). This case would be especially interesting  since it corresponds to outer NS surface where magnetic helicity evolution can be crucial for X-ray bursts observed in magnetars.

\acknowledgments

We are thankful to D.D.~Sokoloff and L.B.~Leinson for helpful discussions, as well as to A.~Chugunov for useful comments. One of the authors (M.D.) is thankful to the Competitiveness Improvement Program at the Tomsk State University and RFBR (Research Project No.~18-02-00149a) for a partial support.

\appendix

\section{Spin distribution functions in a magnetic field
\label{sec:SPINDISTR}}

First we shall study the electron spin distribution function. Let us decompose the energy levels in eq.~\eqref{levels} over a small paramagnetic term, which is equivalent to the consideration of  weak magnetic fields $e H\ll \varepsilon_p^2$. In this case, one has
\begin{equation}
  \mathcal{E}_e(n,p_z,\lambda) \approx \sqrt{m_e^2 + p_z^2 +eH(2n +1)}
  \left[
    1 + \frac{eH\lambda}{2(m_e^2 + p_z^2 +eH(2n +1))}
  \right].
\end{equation}
The distribution function of electrons, accounting for the spin indexes, reads
\begin{equation}\label{eq:distribtot}
  f_{\lambda\lambda'}^{(e)} =
  \frac{\delta_{\lambda\lambda'}}{\exp\{\beta[\mathcal{E}_e(n,p_z,\lambda)-\mu_e]\}+1},
\end{equation}
where $\beta = 1/T$ is the reciprocal temperature.

Then, we recast the exact equilibrium Fermi distribution in eq.~\eqref{eq:distribtot} for electrons in the WKB approximation,\footnote{When $n\gg 1$ or $eH(2n +1)=p_{\perp}^2$, i.e. $\sqrt{m_e^2 + p_z^2 +eH(2n +1)}\Longrightarrow \varepsilon_p=\sqrt{m_e^2 + p^2}$.} separating the paramagnetic term,
\begin{equation}\label{separation}
  f_{\lambda\lambda'}^{(e)} \approx
  \frac{1}{2}
  \left[
    f_\mathrm{eq}^{(e)}\delta_{\lambda\lambda'} +
    \lambda\delta_{\lambda\lambda'}\frac{\mathrm{d}f_\mathrm{eq}^{(e)}}
    {\mathrm{d}\varepsilon_p}\frac{\mu_{B}H}{\gamma}
  \right].
\end{equation}
Here $\varepsilon_p=\sqrt{p^2 + m_e^2}$ and $f_\mathrm{eq}(\varepsilon_p)=g_s[e^{(\varepsilon_p - \mu_e)/T }+ 1]^{-1}$ includes the factor Lande $g_s=2$.

Using the definition of the spin distribution function as
\begin{equation}
  f_{\lambda\lambda'}^{(e)} =
  \frac{1}{2}
  \left[
    f_\mathrm{eq}^{(e)}\delta_{\lambda\lambda'} +
    \left(
      \bm{\sigma}
    \right)_{\lambda\lambda'}\mathbf{S}^{(e)}
  \right],
\end{equation}
and assuming that $ \lambda\delta_{\lambda\lambda'} = \left(\sigma_{z}\right)_{\lambda\lambda'}$, we obtain that $S_{z}^{(e)}=(\mu_{B}H/\gamma) \mathrm{d}f_\mathrm{eq}^{(e)}/\mathrm{d}\varepsilon_p$. Returning to the vector notations, one gets that
\begin{equation}
  \mathbf{S}_\mathrm{eq}^{(e)}=
  \frac{\mu_{B}\mathbf{H}}{\gamma}
  \frac{\mathrm{d}f_\mathrm{eq}^{(e)}}{\mathrm{d}\varepsilon_p},
\end{equation}
which coincides with eq.~\eqref{spin_0}.

To derive the spin distribution function in eq.~\eqref{spin_00} we should choose the appropriate sign of the electric charge in eq.~\eqref{levels} and the opposite sign of $\lambda$. Performing analogous calculations, we obtain  eq.~\eqref{spin_00}.

\section{Spin contribution originated by the pseudoscalar
\label{sec:GAMMA5}}

In ``weak" magnetic fields, $eB\ll \varepsilon_p^2$, we use the WKB approximation with the plane wave decomposition in the Schr\"{o}dinger representation,
\begin{eqnarray}\label{pseudoscalar}
2\mathrm{i}m_e\langle\bar{\psi}\gamma_5\psi\rangle=&&\frac{2\mathrm{i}m_e}{V}\sum_{{\bf p},{\bf p^{\prime}},rr^{\prime}}\frac{1}{\sqrt{4\varepsilon_p\varepsilon_{p^{\prime}}}}\Bigl[ e^{\mathrm{i}({\bf p}^{\prime} - {\bf p}){\bf x}}\langle\hat{b}^+_r(p)\hat{b}_{r^{\prime}}(p^{\prime})\rangle\bar{u}_{r}(p)\gamma_5u_{r^{\prime}}(p^{\prime})\nonumber\\&& - e^{\mathrm{i}({\bf p} - {\bf p}^{\prime}){\bf x}}\langle\hat{d}^+_{r^{\prime}}(p^{\prime})\hat{d}_{r}(p)\rangle\bar{v}_r(p)\gamma_5v_{r^{\prime}}(p^{\prime}) \Bigr],
\end{eqnarray}
where $\varepsilon_p=\sqrt{p^2 + m_e^2}$. The distribution functions in the momentum representation is defined by the mean value $\langle\hat{b}^+_r(p)\hat{b}_{r^{\prime}}(p^{\prime})\rangle=\text{Tr}[\hat{\rho}(t)\hat{b}^+_r(p)\hat{b}_{r^{\prime}}(p^{\prime})]=f^{(e)}_{{\bf p^{\prime}}r^{\prime},{\bf p}r}(t)=f^{(e)}_{{\bf P + \frac{{\bf k}}{2}}r^{\prime},{\bf P - \frac{{\bf k}}{2}}r}(t)$ and $\langle\hat{d}^+_{r^{\prime}}(p^{\prime})\hat{d}_r(p)\rangle=\text{Tr}[\hat{\rho}(t)\hat{d}^+_{r^{\prime}}(p^{\prime})\hat{d}_{r}(p)]=f^{(\bar{e})}_{{\bf p}r,{\bf p^{\prime}}r^{\prime}}(t)=f^{(\bar{e})}_{{\bf P - \frac{{\bf k}}{2}}r,{\bf P + \frac{{\bf k}}{2}}r^{\prime}}(t)$. These functions are given, in general, by the non-equilibrium statistical operator $\hat{\rho}(t)$; cf. ref.~\cite{Peletminsky}. Here $2{\bf P}= {\bf p}^{\prime} + {\bf p}$, ${\bf k}= {\bf p}^{\prime} - {\bf p}$ is the momentum transfer.  We remind the definition of the Wigner distribution functions:
\begin{eqnarray}\label{Wigner}
&&\sum_{\bf k}e^{i{\bf kx}}f^{(e)}_{{\bf P} + \frac{{\bf k}}{2}r^{\prime},{\bf P} - \frac{{\bf k}}{2}r}(t)=f^{(e)}({\bf P}, {\bf x},t)\frac{\delta_{r^{\prime}r}}{2} + {\bf S}^{(e)}({\bf P}, {\bf x},t)\frac{(\bm{\sigma})_{r^{\prime}r}}{2},\nonumber\\&&
\sum_{\bf k}e^{-i{\bf kx}}f^{(\bar{e})}_{{\bf P} - \frac{{\bf k}}{2}r,{\bf P} + \frac{{\bf k}}{2}r^{\prime}}(t)=f^{(\bar{e})}({\bf P}, {\bf x},t)\frac{\delta_{r^{\prime}r}}{2} + {\bf S}^{(\bar{e})}({\bf P}, {\bf x},t)\frac{(\bm{\sigma})_{rr^{\prime}}}{2}.
\end{eqnarray}
Note that the sequence of spin indexes is important in Wigner's spin distribution terms in eq.~(\ref{Wigner}): it is different in electron and positron cases.

In what follows we use the WKB approximation, ${\bf k}\ll {\bf P}$, that corresponds to a low inhomogeneous medium. We provide a useful relation,
\begin{equation}
\varepsilon_{{\bf P}\pm \frac{{\bf k}}{2}}=\sqrt{m_e^2 + P^2 \pm {\bf Pk} + k^2/2}=\varepsilon_P\left[1 \pm \frac{{\bf Pk}}{2\varepsilon^2_P} + \mathcal{O}(k^2)\right]
\end{equation}
which results in the normalization factor $1/\sqrt{4\varepsilon_{p^{\prime}}\varepsilon_p}=
1/2\varepsilon_P[1 +  \mathcal{O}(k^2)]=1/2\varepsilon_P $. Here $\varepsilon_P=\sqrt{P^2 + m_e^2}$ in eq.~(\ref{pseudoscalar}).

The bispinors for a free particle have the form,
\begin{equation}\label{col}
  u_{r^{\prime}}
  \left(
    {\bf P} + \frac{{\bf k}}{2}
  \right) =
  \sqrt{\varepsilon_P
  \left(
    1 + {\bf Pk}/2\varepsilon_P^2
  \right) + 
  m_e}
  \left(
    \begin{array}{c}
      {\rm I}
      \\
      \dfrac{(\bm{\sigma}\cdot[{\bf P} +{\bf k}/2])}{\varepsilon_{{\bf P} +{\bf k}/2} + m_e}
    \end{array}
  \right)
  \otimes
  \varphi_{r^{\prime}}.
\end{equation}
Thus one gets
\begin{equation}\label{col1}
  \gamma_5u_{r^{\prime}}
  \left(
    {\bf P} + \frac{{\bf k}}{2}
  \right)=\sqrt{\varepsilon_P
  \left(
    1 + {\bf Pk}/2\varepsilon_P^2
  \right) + 
  m_e}
  \left(
    \begin{array}{c}
      \dfrac{(\bm{\sigma}\cdot[{\bf P} +{\bf k}/2])}{\varepsilon_{{\bf P} +{\bf k}/2} + m_e}
      \\
      {\rm I}
    \end{array}
  \right)
  \otimes
  \varphi_{r^{\prime}}.
\end{equation}
In eqs.~\eqref{col} and~\eqref{col1}, {\rm I} is the unit $2\times 2$ matrix.

Analogously we can obtain that
\begin{equation}\label{line}
\bar{u}_r\left({\bf P} - \frac{{\bf k}}{2}\right)=\sqrt{\varepsilon_P\left(1 - {\bf Pk}/2\varepsilon_P^2\right) + m_e}\varphi^\dag_r
\otimes
\left({\rm I}, -\frac{\bm{\sigma}({\bf P} -{\bf k}/2)}{\varepsilon_{{\bf P} -{\bf k}/2} + m_e}\right).
\end{equation}
Using eqs.~\eqref{col}-\eqref{line}, we get the electron matrix element in eq.~(\ref{pseudoscalar}) as
\begin{multline}\label{electron-element}
\frac{1}{2\sqrt{\varepsilon_{{\bf P} + {\bf k}/2}\varepsilon_{{\bf P} - {\bf k}/2}}}\bar{u}_r\left({\bf P} - \frac{{\bf k}}{2}\right)\gamma_5u_{r^{\prime}}\left({\bf P} + \frac{{\bf k}}{2}\right)=\left(\frac{\varepsilon_P + m_e + O(k^2)}{2\varepsilon_P (1 + O(k^2))}\right)
\\
\times\varphi^\dagger_r\left[\frac{(\bm{\sigma}{\bf P} + \bm{\sigma}{\bf k}/2)}{(\varepsilon_P + m_e)[1 + {\bf Pk}/2\varepsilon_P(\varepsilon_P + m_e)]} -\frac{(\bm{\sigma}{\bf P} - \bm{\sigma}{\bf k}/2)}{(\varepsilon_P + m_e)[1 - {\bf Pk}/2\varepsilon_P(\varepsilon_P + m_e)]}\right]\varphi_{r^{\prime}}
\\
=\frac{1}{2\varepsilon_P}\left[\left(1 - \frac{{\bf Pk}}{2\varepsilon_P(\varepsilon_P + m_e)}\right)\Bigl(\bm{\sigma}_{rr^{\prime}}({\bf P} + {\bf k}/2)\Bigr) - \left(1 + \frac{{\bf Pk}}{2\varepsilon_P(\varepsilon_P + m_e)}\right)\Bigl(\bm{\sigma}_{rr^{\prime}}({\bf P} - {\bf k}/2)\Bigr)\right]
\\
=\frac{1}{2\varepsilon_P}(\sigma_j)_{rr^{\prime}}\left(k_j -\frac{P_j({\bf Pk})}{\varepsilon_P(\varepsilon_P + m_e)}\right) + \mathcal{O}(k^2).
\end{multline}

For positrons we use charge conjugation $v_r(p)=U_\mathrm{C}\bar{u}^\mathrm{T}_r(p)$, where $U_\mathrm{C}=\mathrm{i}\gamma^2\gamma^0$, that gives the corresponding positron matrix element entering eq.~(\ref{pseudoscalar}) and resulting from the electron one in eq.~(\ref{electron-element}),
\begin{multline}\label{positron-element}
\frac{1}{2\varepsilon_P}\bar{v}_r({\bf P} - {\bf k}/2)\gamma_5v_{r^{\prime}}({\bf P} + {\bf k}/2)= - \frac{1}{2\varepsilon_P}\bar{u}_{r^{\prime}}({\bf P} + {\bf k}/2)\gamma_5u_r({\bf P} - {\bf k}/2)
\\
= - \frac{1}{2\varepsilon_P}(\sigma_j)_{r^{\prime}r}\left( - k_j +\frac{P_j({\bf Pk})}{\varepsilon_P(\varepsilon_P + m_e)}\right)=\frac{1}{2\varepsilon_P}(\sigma_j)_{r^{\prime}r}\left( k_j -\frac{P_j({\bf Pk})}{\varepsilon_P(\varepsilon_P + m_e)}\right).
\end{multline}
One can see that the difference of two matrix elements in eqs.~(\ref{electron-element}) and~(\ref{positron-element}), concerns  the different sequence of spin indexes only.

Then accounting for the definition of Wigner functions (\ref{Wigner}),
\begin{align}
  \sum_{\bf k}
  \mathrm{i}k_je^{i{\bf kx}}f^{(e)}_{{\bf P}+{\bf k}/2 r^{\prime} {\bf P}-{\bf k}/2 r}(t)
  = &
  \frac{\partial}{\partial x^j}f^{(e)}_{r^{\prime}r}({\bf P},{\bf x},t)
  \notag
  \\
  & =
  \frac{\partial f^{(e)}({\bf P},{\bf x},t)}{\partial x^j}\frac{\delta_{r^{\prime}r}}{2} +
  \frac{\partial S^{(e)}({\bf P},{\bf x},t)}{\partial x^j}\frac{\sigma_{r^{\prime}r}}{2},
\end{align}
changing ${\bf k}\to -{\bf k}$ in eq.~(\ref{Wigner}) for positrons, while retaining the sequence of spin indexes,
\begin{align}
  \sum_{\bf k}
  \mathrm{i}k_je^{i{\bf kx}}f^{(\bar{e})}_{{\bf P}+{\bf k}/2 r {\bf P}-{\bf k}/2 r^{\prime}}(t)
  = &
  \frac{\partial}{\partial x^j}f^{(\bar{e})}_{rr^{\prime}}({\bf P},{\bf x},t)
  \notag
  \\
  & =
  \frac{\partial f^{(\bar{e})}({\bf P},{\bf x},t)}{\partial x^j}\frac{\delta_{r^{\prime}r}}{2} +
  \frac{\partial S^{(\bar{e})}({\bf P},{\bf x},t)}{\partial x^j}\frac{\sigma_{rr^{\prime}}}{2},
\end{align}
then changing the remaining sum $V^{-1}\sum_{{\bf P}}\to \smallint \mathrm{d}^3P/(2\pi)^3$, one can find the mean pseudoscalar we are looking for,
\begin{align}\label{pseudoscalar2}
  2\mathrm{i}m_e\langle\bar{\psi}\gamma_5\psi\rangle = &
  2m_e\int \frac{\mathrm{d}^3P}{(2\pi)^3}\sum_{r,r^{\prime}}\frac{1}{2\varepsilon_P}   
  \nonumber
  \displaybreak[1]
  \\
  & \times
  \Bigg[\left(\frac{\delta_{r^{\prime}r}}{2}\frac{\partial f^{(e)}({\bf P},{\bf x},t)}{\partial x^j} + \frac{(\sigma_k)_{r^{\prime}r}}{2}\frac{\partial S_k^{(e)}({\bf P},{\bf x},t)}{\partial x^j}\right)(\sigma_j)_{rr^{\prime}}
  \nonumber
  \displaybreak[1]
  \\
  & -
  \frac{P_jP_m}{\varepsilon_P(\varepsilon_P + m_e)}\left(\frac{\delta_{r^{\prime}r}}{2}\frac{\partial f^{(e)}({\bf P},{\bf x},t)}{\partial x^m} + \frac{(\sigma_k)_{r^{\prime}r}}{2}\frac{\partial S_k^{(e)}({\bf P},{\bf x},t)}{\partial x^m}\right)(\sigma_j)_{rr^{\prime}}
  \nonumber
  \displaybreak[1]
  \\
  & -
  \left(\frac{\delta_{r^{\prime}r}}{2}\frac{\partial f^{(\bar{e})}({\bf P},{\bf x},t)}{\partial x^j} + \frac{(\sigma_k)_{rr^{\prime}}}{2}\frac{\partial S_k^{(\bar{e})}({\bf P},{\bf x},t)}{\partial x^j}\right)(\sigma_j)_{r^{\prime}r}
  \nonumber
  \displaybreak[1]
  \\
  & +
  \frac{P_jP_m}{\varepsilon_P(\varepsilon_P + m_e)}\left(\frac{\delta_{r^{\prime}r}}{2}\frac{\partial f^{(\bar{e})}({\bf P},{\bf x},t)}{\partial x^m} + \frac{(\sigma_k)_{rr^{\prime}}}{2}\frac{\partial S_k^{(\bar{e})}({\bf P},{\bf x},t)}{\partial x^m}\right)(\sigma_j)_{r^{\prime}r}\Bigg].
\end{align}

One can easily see that, summing in eq.~(\ref{pseudoscalar2}) over spin projections, only spin distributions contribute to the mean pseudoscalar,
\begin{align}\label{main}
2\mathrm{i}m_e\langle\bar{\psi}\gamma_5\psi\rangle=&2m_e\int \frac{\mathrm{d}^3P}{2\varepsilon_P(2\pi)^3}\Bigg[\frac{\partial S_j^{(e)}({\bf P},{\bf x},t)}{\partial x^j} - \frac{\partial S_j^{(\bar{e})}({\bf P},{\bf x},t)}{\partial x^j}
\nonumber
\\
& -\frac{P_jP_m}{\varepsilon_P(\varepsilon_P + m_e)}\left(\frac{\partial S_j^{(e)}({\bf P},{\bf x},t)}{\partial x^m} - \frac{\partial S_j^{(\bar{e})}({\bf P},{\bf x},t)}{\partial x^m}\right)\Bigg].
\end{align}
Integrating eq.~(\ref{main}) over space, $V^{-1}\int \mathrm{d}^3x$, to get a violation term in eq.~(\ref{magnetization}), and marking that in the last line in eq.~(\ref{main}), the symmetric part of the tensor $1/2\big(\partial S_j^{(e,\bar{e})}/\partial x^m + \partial S_m^{(e,\bar{e})}/\partial x^j\big)$ contributes only, one can write that last line for both electrons and positrons as
\begin{multline}
-2m_e\int \frac{\mathrm{d}^3P}{2\varepsilon_P(2\pi)^3}\frac{P_jP_m}{\varepsilon_P(\varepsilon_P + m_e)}\int \frac{\mathrm{d}^3x}{V}\frac{1}{2}\left[\frac{\partial S_j^{(e,\bar{e})}({\bf P},{\bf x},t)}{\partial x^m} + \frac{\partial S_m^{(e,\bar{e})}({\bf P},{\bf x},t)}{\partial x^j}\right]
\\
=-\frac{1}{3}\int \frac{\mathrm{d}^3x}{V}\int \frac{\mathrm{d}^3P}{\gamma(2\pi)^3}\frac{P^2}{\varepsilon_p(\varepsilon_P + m_e)}\frac{\partial S_j^{(e,\bar{e})}({\bf P},{\bf x},t)}{\partial x^j}.
\end{multline}
Thus the new violation term in eq.~(\ref{new_law}) takes the form,
\begin{align}
2\mathrm{i}m_e\int \frac{\mathrm{d}^3x}{V}\langle \bar{\psi}\gamma_5\psi\rangle=&\int\frac{\mathrm{d}^3x}{V}\int\frac{\mathrm{d}^3P}{\gamma(2\pi)^3}
\nonumber
\\
& \times
\left[\left(\frac{\partial S_j^{(e)}({\bf P},{\bf x},t)}{\partial x^j} - \frac{\partial S_j^{(\bar{e})}({\bf P},{\bf x},t)}{\partial x^j}\right)\Bigl[ 1 -\frac{1}{3}\left(1 - \frac{1}{\gamma}\right)\Bigr]\right]
\nonumber
\\
& =
- \oint_S(\bm{\mathcal{S}}_5\cdot{\bf n})\frac{\mathrm{d}^2S}{V},
\end{align}
where the pseudovector
\begin{equation}\label{magnetization3}
  \bm{\mathcal{S}}_5=-\int\frac{\mathrm{d}^3P}{\gamma(2\pi)^3}
  \left(
    {\bf S}^{(e)}({\bf P},{\bf x},t) - {\bf S}^{(\bar{e})}({\bf P},{\bf x},t)
  \right)
  \left(
    \frac{2}{3} + \frac{1}{3\gamma}
  \right),
\end{equation}
reproduces that in eq.~(\ref{magnetization}).

\end{document}